# Experimental Evaluation of Server Centric Passive Optical Network Based Data Centre Architecture

Azza E. A. Eltraify, Mohamed O. I. Musa, Ahmed Al-Quzweeni and Jaafar M.H. Elmirghani
*School of Electronic and Electrical Engineering, University of Leeds, Leeds, LS2 9JT, UK*
E-mail: {scaeae, m.musa, ml13anaq, j.m.h.elmirghani}@leeds.ac.uk

**ABSTRACT**
Passive optical networks (PON) technology has recently been proposed as a solution for scalability, energy efficiency, high capacity, low cost, flexibility and oversubscription issues in data centres. This paper experimentally demonstrates and discusses the implementation of a server centric PON based data centre architecture with high speed and reliability. The architecture is set up using a set of servers grouped into racks directly connected together and to the Optical Line Terminal (OLT) through gateway servers. The switching and routing functionalities have been embedded into servers using 4x10GE Xilinx NetFPGA. Flow continuity has been observed through live video streaming using IP cameras transmitting over up to 110 km optical connections through WDM nodes and the PON network.
**Keywords**: Passive optical Networks, NetFPGA, ONU, PON, NIC, WDM

## 1. INTRODUCTION

In recent years, higher data rates and power efficiency in data centres have become essential, hence several studies have been carried out in optimising energy efficiency and architectures in data centres and the core networks connecting them [1]-[10]. In order to validate the results of these studies, experimental work was needed to deploy and evaluate these architectures.

As shown in Figure 1 one of our previously designed PON based data centre network architectures [11] proposed a server centric architecture that provides high speed communication between servers. The architecture subdivides the data centre into PON cells, where each cell contains servers grouped in racks with an OLT providing control and interfaces. To facilitate the communication between servers in different groups within the same rack an optical backplane is used [12]. Each rack has a group of servers that handles the aggregation of traffic between the rack and the OLT. Other groups are directly connected to different groups within other racks to provide inter-rack communication. Data between the OLT and the racks is routed through a coupler (Time Division Multiplexing (TDM)) or via an arrayed Waveguide Grating (AWGR) that makes use of Wavelength Division Multiplexing (WDM) [11], [13] - [15].

Studies have been carried out to find a solution for the problem of higher data rate demands within data centres, as well as faster processing, energy efficiency and scalability. One of these studies introduced the NetFPGA board, a Field Programmable Gate Arrays (FPGA) based networking solution. FPGAs were developed to enhance switching, routing and processing of network data. The NetFPGA Platform is an open source hardware and software platform composed of a large programmable Xilinx FPGA, PCI interface, static RAMs (SRAMs), Double Data Rate (DDR2) SDRAM, quad-port physical-layer transceiver (PHY) for transmission and reception [16]. In this paper the NetFPGA-10G is used as a Network Interface Card (NIC) for the servers in the processing cells. It is an FPGA-based PCI Express board with four 10-Gigabit SFP+ interfaces, an x8 gen1 PCIe adapter card incorporating Xilinx's Virtex-5 TX240TFPGA [16].

In this paper an experimental implementation of the proposed data centre architecture has been carried out to test its operation when two instances of the architecture are connected through different platforms over long distances of over 110 km of optical interconnections passing through a core network and a PON cell.

## 2. ARCHITECTURE IMPLEMENTATION

The implementation of the server centric passive optical network data centre architecture shown in Figure 1 was achieved by using three racks, each containing three servers connected via a 10 Gbps Cisco Switch. Racks are connected together using an optical link via a media converter. Figure 2 Shows the IP addressing and connectivity of the servers within the same rack and among different racks. Figure 3 shows the experimental deployment of the architecture in our laboratories. As a solution to the unavailability of the optical backplane an electronic switch was used to connect servers and the servers operating system handled the routing and relaying of traffic.

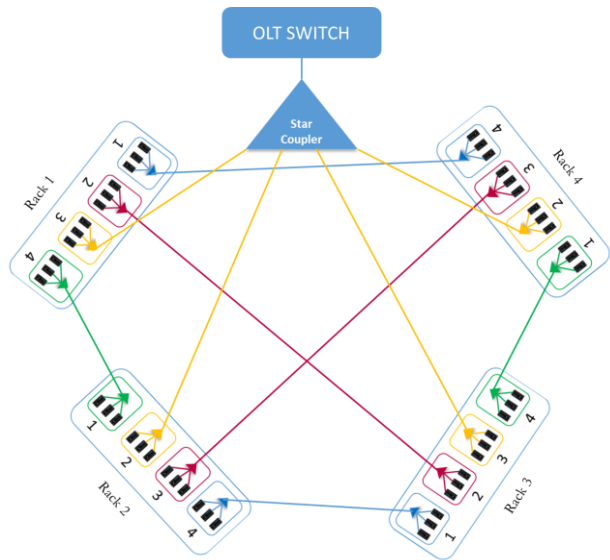

*Figure 1. Server Centric PON based data centre network architecture [17]*

To implement the routing functionalities relay servers were equipped with MikroTik OS; a specialized Linux Kernel operating system, while the other servers operated using Ubuntu Operating system.

The server centric PON architecture is utilised as a processing cell. Each rack is given a different IP network and racks communicate through gateways among themselves. The variety of connections and alternative routes between servers provides high reliability. Figure 2 demonstrates the networks and IP assignment and it shows the gateway servers within each rack that are used for traffic relaying. Relay servers are configured with two IP addresses for intra-rack communication and inter-rack communication. All other servers in the network were set up with the gateway IP address for inter-rack communication, [19].

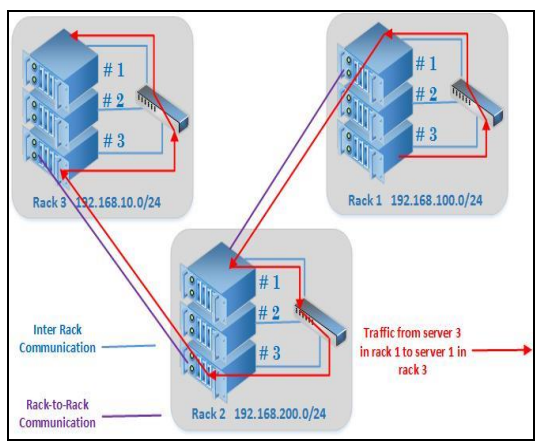

*Figure 2. Intra-rack and inter-rack communication in the proposed PON based data centre architecture*

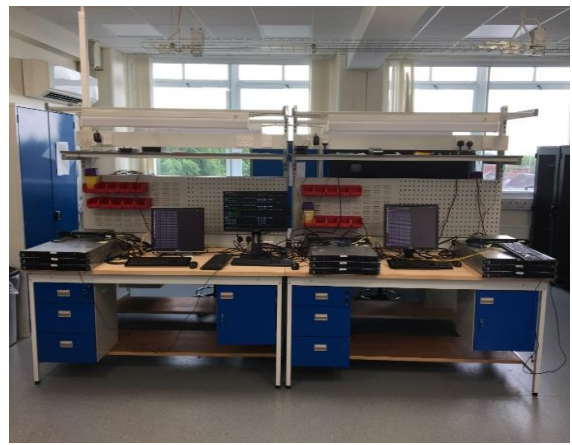

*Figure 3. Server Centric Architecture Experimental Implementation*

An IP over DWDM core network is used to demonstrate communication between multiple data centres located in multiple cities with over 100 km of optical fibre links as shown in Figure 4 [20]-[29]. Each core node is 100 Gbps MRV/ADVA DWDM node. Multiplexers/Demultiplexers are used on the C-band supporting 80 wavelengths, and EDFAs and dispersion compensators are used to improve the quality of the optical signal over the long distances considered. This setup gave the ability to test applications, traffic flow and latency [30]-38] over various long distances similar to actual distances between cities in a typical core network [18].

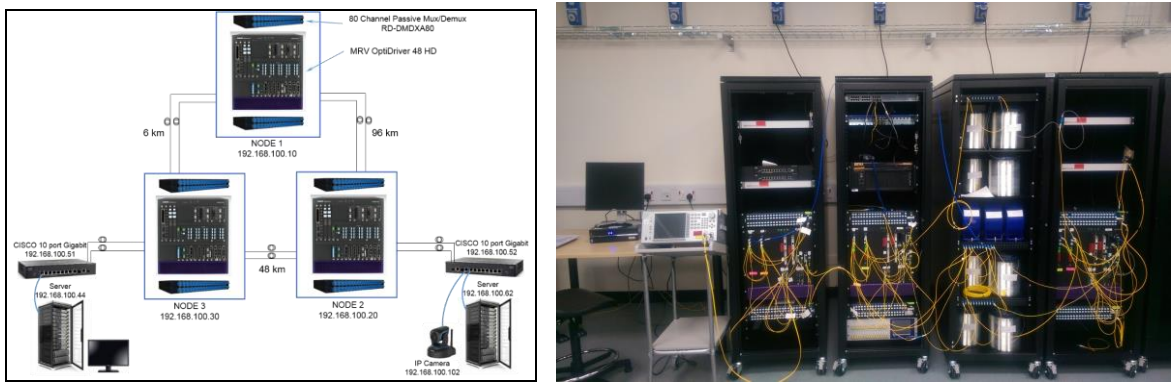

*Figure 4. IP/WDM Core Network Nodes*

In order to demonstrate end-to-end communication, two processing nodes were connected through an IP/WDM core network and a PON cell as shown in Figure 5, an IPTV Camera was connected to the first processing cell and streams live video throughout the processing cell, over the IP/WDM Core networks nodes, and to the OLT/ONU.

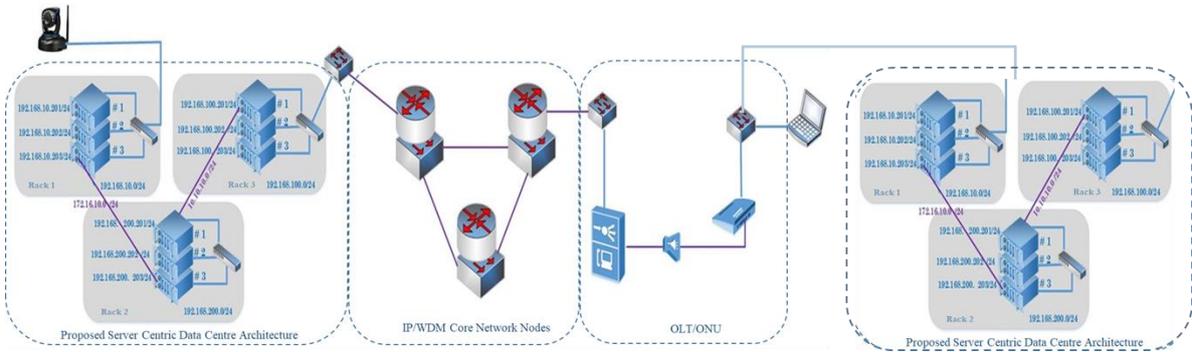

*Figure 5. The implemented end-to-end System of the Proposed Server Centric Data Centre Architecture, IP/WDM Core Network Nodes and OLT/ONU*

## 3. RESULTS

The end-to end system connectivity was accomplished by connecting two processing nodes, an IP/WDM Core Network Nodes, and an OLT/ optical network unit (ONU). High quality real time video streaming using an IoT camera and an IPTV camera over a distance of over 110 km demonstrated continuous flow of traffic.

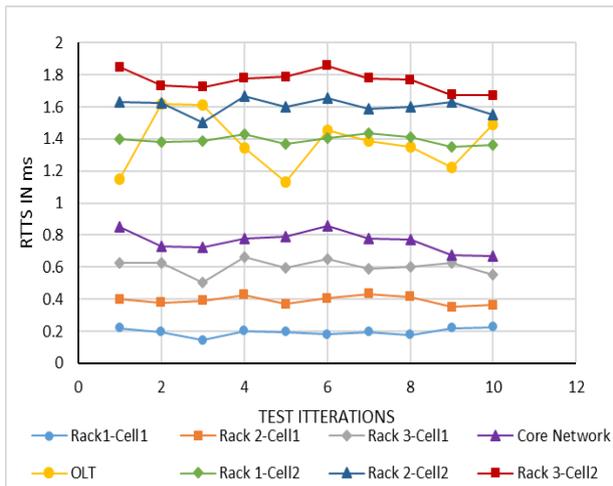
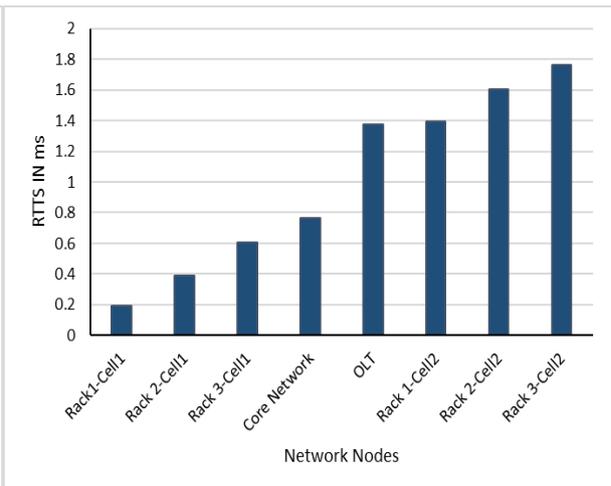

*Figure 6. The round-trip time of all node*  *Figure 7. Average RTT for all nodes*

The ICMP protocol and route tracing were used to ensure proper routing and configuration of the network. ICMP signals were sent from servers on the all the nodes within the network with successful replies in every test resulting in 0% packet loss. A traceroute signal was sent from the first server in the processing cell to the last server in the other processing cell at the end of the system setup. The latency at each point was measured by obtaining the round-trip time (RTT) of each hop. The traceroute signal was sent 10 times, each with 150 results. Figure 6 shows the 10 test iterations and the results of each hop's average RTT.

The end-to-end system has proven to have low latency, with RTT less than 2 ms (ranging from 0.1958 ms to 1.7619 ms) as shown in Figure 6, despite the traffic being relayed over long distances and diverse platforms. Figure 7 shows that there is on average a 0.7 ms delay between the core network and OLT while all other nodes maintained an average of 0.2 ms delay which is due to the number of nodes within the core network.

## 4. CONCLUSIONS

This paper provided an experimental evaluation of the performance of a proposed Service Centric PON based data centre architecture that is cost and power efficient. The performance was illustrated by using ICMP signals and video streaming. The NetFPGA boards for NIC communication facilitated the optical interconnections between servers while maintaining high speed and high data rate. In our previous experiment [19], 5 nodes were tested for latency, whereas in this paper 8 nodes were tested and the results showed that the increase in delay was insignificant when expanding the number of nodes at the data centre level.

### ACKNOWLEDGEMENTS

The authors would like to acknowledge funding from the Engineering and Physical Sciences Research Council (EPSRC), INTERNET (EP/H040536/1) and STAR (EP/K016873/1) and from the EPSRC TOWS (EP/S016570/1) project. All data are provided in full in the results section of this paper.

### REFERENCES

[1] X. Dong, T. El-Gorashi, and J. M. Elmirghani, "IP over WDM networks employing renewable energy sources," *J. Lightwave Technol.,* vol. 29, no. 1, pp. 3–14, 2011.

[2] X. Dong, T. El-Gorashi, and J. M. H. Elmirghani, "Green IP over WDM networks with data centers," *J. Lightwave Technol.,* vol. 29, no. 12, pp. 1861–1880, June 2011.

[3] X. Dong, T. E. H. El-Gorashi, and J. M. H. Elmirghani, "On the energy efficiency of physical topology design for IP over WDM networks," *J. Lightwave Technol.,* vol. 30, no. 12, pp. 1931–1936, 2012.

[4] A. Lawey, T. El-Gorashi, and J. Elmirghani, "Distributed energy efficient clouds over core networks," *J. Lightwave Technol.,* vol. 32, no. 7, pp. 1261–1281, Apr. 2014.

[5] N. Osman, T. El-Gorashi, L. Krug, and J. Elmirghani, "Energy-efficient future high-definition TV," *J. Lightwave Technol.,* vol. 32, no. 13, pp. 2364–2381, July 2014.

[6] A. Lawey, T. El-Gorashi, and J. Elmirghani, "BitTorrent content distribution in optical networks," *J. Lightwave Technol.,* vol. 32, no. 21, pp. 4209–4225, Nov. 2014.

[7] L. Nonde, T. El-Gorashi, and J. Elmirghani, "Energy efficient virtual network embedding for cloud networks," *J. Lightwave Technol.,* vol. 33, no. 9, pp. 1828–1849, May 2015.

[8] M. Musa, T. Elgorashi, and J. Elmirghani, "Energy efficient survivable IP-over-WDM networks with network coding," *J. Opt. Commun. Netw.,* vol. 9, no. 3, pp. 207–217, 2017.

[9] Musa, Mohamed, Taisir Elgorashi, and Jaafar Elmirghani. "Bounds for Energy-Efficient Survivable IP Over WDM Networks With Network Coding" *Journal of Optical Communications and Networking* 10.5 (2018): 471-481, 2018.

[10] Elmirghani, J. M. H., T. Klein, K. Hinton, L. Nonde, A. Q. Lawey, T. E. H. El-Gorashi, M. O. I. Musa, and X. Dong. "GreenTouch GreenMeter core network energy-efficiency improvement measures and optimization" *Journal of Optical Communications and Networking* 10, no. 2 (2018): A250-A269.

[11] Ali Abdullah Hammadi: "Future PON Data Centre Networks", *University of Leeds, School of Electronic and Electrical Engineering*, Aug. 2016.

[12] J. Beals IV, N. Bamiedakis, A. Wonfor, R. Penty, I. White, J. DeGroot Jr, et al.: "A terabit capacity passive polymer optical backplane based on a novel meshed waveguide architecture", *Applied Physics A,* vol. 95, pp. 983-988, 2009.

[13] Hammadi, Ali, Taisir EH El-Gorashi, Mohamed OI Musa, and Jaafar MH Elmirghani, "Server-centric PON data center architecture" *In Transparent Optical Networks (ICTON), 2016 18th International Conference on,* pp. 1-4. IEEE, 2016.

[14] A. Hammadi, Mohamed O. I. Musa, T. E. H. El-Gorashi, and J.M.H. Elmirghani, "Resource Provisioning for Cloud PON AWGR-Based Data Center Architecture", *21st European Conference on Network and Optical Communications (NOC),* Portugal, 2016.

[15] Hammadi, Ali, Taisir EH El-Gorashi, and Jaafar MH Elmirghani. "High performance AWGR PONs in data centre networks", *In Transparent Optical Networks (ICTON), 2015 17th International Conference on,* pp. 1-5. IEEE, 2015.

[16] John W. Lockwood, Nick McKeown, Greg Watson, Glen Gibb, Paul Hartke, Jad Naous, Ramanan Raghuraman, and Jianying Luo; "NetFPGA - An Open Platform for Gigabit-rate Network Switching and Routing"; *MSE 2007,* San Diego, June 2007.

[17] A. Hammadi, T. E. El-Gorashi, and J. M. H. Elmirghani; "PONs in Future Cloud Data Centers"; *IEEE Communications Magazine.*

[18] X. Dong, T. El-Gorashi, J.M.H. Elmirghani, "Green IP Over WDM Networks With Data Centers", *IEEE Journal of Lightwave Technology,* vol.29, no.12, pp.1861-1880, June 2011.

[19] Eltraify, Azza EA, Mohamed OI Musa, Ahmed Al-Quzweeni, and Jaafar MH Elmirghani. "Experimental Evaluation of Passive Optical Network Based Data Centre Architecture." In *2018 20th International Conference on Transparent Optical Networks (ICTON)*, pp. 1-4. IEEE, 2018.

[20] J. M. H. Elmirghani, L. Nonde, A. Q. Lawey, T. E. H. El-Gorashi, M. O. I. Musa, X. Dong, K. Hinton, and T. Klein, "Energy efficiency measures for future core networks," in 2017 Optical Fiber Communications Conference and Exhibition (OFC), March 2017, pp. 1-3.

[21] H.M.M., Ali, A.Q. Lawey, T.E.H. El-Gorashi, and J.M.H. Elmirghani, "*Future Energy Efficient Data Centers With Disaggregated Servers,*" IEEE/OSA Journal of Lightwave Technology, vol. 35, No. 24, pp. 5361 – 5380, 2017.

[22] B. Bathula, M. Alresheedi, and J.M.H. Elmirghani, "*Energy efficient architectures for optical networks*," Proc IEEE London Communications Symposium, London, Sept. 2009.

[23] M. Musa, T.E.H. El-Gorashi and J.M.H. Elmirghani, "*Bounds for Energy-Efficient Survivable IP Over WDM Networks with Network Coding*," IEEE/OSA Journal of Optical Communications and Networking, vol. 10, no. 5, pp. 471-481, 2018.

[24] B. Bathula, and J.M.H. Elmirghani, "*Energy Efficient Optical Burst Switched (OBS) Networks*," IEEE GLOBECOM'09, Honolulu, Hawaii, USA, November 30-December 04, 2009.

[25] X. Dong, T.E.H. El-Gorashi and J.M.H. Elmirghani, "*Green Optical OFDM Networks,*" IET Optoelectronics, vol. 8, No. 3, pp. 137 – 148, 2014.

[26] M. Musa, T.E.H. El-Gorashi and J.M.H. Elmirghani, "*Energy Efficient Survivable IP-Over-WDM Networks With Network Coding*," IEEE/OSA Journal of Optical Communications and Networking, vol. 9, No. 3, pp. 207-217, 2017.

[27] A. Lawey, T.E.H. El-Gorashi, and J.M.H. Elmirghani, "*BitTorrent Content Distribution in Optical Networks,*" IEEE/OSA Journal of Lightwave Technology, vol. 32, No. 21, pp. 3607 – 3623, 2014.

[28] A.M. Al-Salim, A. Lawey, T.E.H. El-Gorashi, and J.M.H. Elmirghani, "*Energy Efficient Big Data Networks: Impact of Volume and Variety,*" IEEE Transactions on Network and Service Management, vol. 15, No. 1, pp. 458 - 474, 2018.

[29] A.M. Al-Salim, A. Lawey, T.E.H. El-Gorashi, and J.M.H. Elmirghani, "Greening big data networks: velocity impact," *IET Optoelectronics,* vol. 12, No. 3, pp. 126-135, 2018.

[30] N. I. Osman and T. El-Gorashi and J. M. H. Elmirghani, "The impact of content popularity distribution on energy efficient caching," in *2013 15th International Conference on Transparent Optical Networks (ICTON)*, June 2013, pp. 1-6.

[31] N. I. Osman, T. El-Gorashi, L. Krug, and J. M. H. Elmirghani, "Energy Efficient Future High-Definition TV," *Journal of Lightwave Technology, vol. 32, no. 13, pp. 2364-2381,* July 2014.

[32] L. Nonde, T. E. H. El-Gorashi, and J. M. H. Elmirghani, "Virtual Network Embedding Employing Renewable Energy Sources," *in 2016 IEEE Global Communications Conference (GLOBECOM*), Dec 2016, pp. 1-6.

[33] A. Q. Lawey, T. E. H. El-Gorashi, and J. M. H. Elmirghani, "Renewable energy in distributed energy efficient content delivery clouds," in *2015 IEEE International Conference on Communications (ICC),* June 2015, pp. 128-134.

[34] A.N. Al-Quzweeni, A. Lawey, T.E.H. El-Gorashi, and J.M.H. Elmirghani, "Optimized Energy Aware 5G Network Function Virtualization," *IEEE Access*, vol. 7, 2019.

[35] M.S. Hadi, A. Lawey, T.E.H. El-Gorashi, and J.M.H. Elmirghani, "Patient-Centric Cellular Networks Optimization using Big Data Analytics," *IEEE Access*, vol. 7, 2019.

[36] M.S. Hadi, A. Lawey, T.E.H. El-Gorashi, and J.M.H. Elmirghani, "Big Data Analytics for Wireless and Wired Network Design: A Survey, *Elsevier Computer Networks,* vol. 132, No. 2, pp. 180-199, 2018.

[37] S. Igder, S. Bhattacharya, and J. M. H. Elmirghani, "Energy Efficient Fog Servers for Internet of Things Information Piece Delivery (IoTIPD) in a Smart City Vehicular Environment," in *2016 10th International Conference on Next Generation Mobile Applications, Security and Technologies (NGMAST),* Aug 2016, pp. 99-104.

[38] S. H. Mohamed, T. E. H. El-Gorashi, and J. M. H. Elmirghani, "Energy Efficiency of Server-Centric PON Data Center Architecture for Fog Computing," in *2018 20th International Conference on Transparent Optical Networks (ICTON),* July *2018*, pp. 1-4.